\documentclass[letterpaper,twocolumn,english,amssymb,aps,prb]{revtex4}
\usepackage[T1]{fontenc}
\usepackage[latin9]{inputenc}
\usepackage{units}
\usepackage{textcomp}
\usepackage{amsmath}
\usepackage{graphicx}
\usepackage{amssymb}
\usepackage{esint}

\makeatletter
\@ifundefined{textcolor}{}
{%
 \definecolor{BLACK}{gray}{0}
 \definecolor{WHITE}{gray}{1}
 \definecolor{RED}{rgb}{1,0,0}
 \definecolor{GREEN}{rgb}{0,1,0}
 \definecolor{BLUE}{rgb}{0,0,1}
 \definecolor{CYAN}{cmyk}{1,0,0,0}
 \definecolor{MAGENTA}{cmyk}{0,1,0,0}
 \definecolor{YELLOW}{cmyk}{0,0,1,0}
 }

\makeatother

\usepackage{babel}

\begin{document}

\title{Perfect Andreev Reflection of Helical Edge Modes in Inverted InAs/GaSb
Quantum Wells}

\author{Ivan Knez}

\affiliation{Department of Physics and Astronomy, Rice University, Houston, TX
77251-1892}

\author{Rui-Rui Du}

\affiliation{Department of Physics and Astronomy, Rice University, Houston, TX
77251-1892}

\author{Gerard Sullivan}

\affiliation{Teledyne Scientific and Imaging, Thousand Oaks, CA 91630}

\maketitle


\textbf{Quantum Spin Hall Insulator (QSHI) is a two-dimensional variant
of a novel class of materials characterized by topological order,
whose unique properties have recently triggered much interest and
excitement in the condensed matter community.\citep{1,2} Most notably,
topological properties of these systems hold great promise in mitigating
the difficult problem of decoherence in implementations of quantum
computers.\citep{3} Although QSHI has been theoretically predicted
in a few different materials,\citep{4,5,6,7} so far only the semiconductor
systems of HgTe/CdTe\citep{8} and, more recently, inverted InAs/GaSb,\citep{9}
have shown direct evidence for the existence of this phase. Ideally
insulating in the bulk, QSHI is characterized by one-dimensional channels
at the sample perimeter, which have helical property, with carrier
spin tied to the carrier direction of motion, and protected from back-scattering
by time-reversal symmetry. Here we experimentally show that QSH edge
channels in InAs/GaSb exhibit perfect Andreev reflection (AR), validating
their helical property and topological protection from back-scattering. }

\textcompwordmark{}

Much of the transport phenomenology of QSHI has been established in
a set of remarkable experiments in HgTe material system,\citep{8,10}
including the quantized conductance and the non-local character of
the QSH edge modes. Combining QSHI with superconductors is the next
experimental challenge, posing fundamental questions regarding the
nature of topological superconductors and the possible realizations
of Majorana fermion excitations.\citep{3,11,12,13} Recently it has
been theoretically suggested that Andreev reflection can be used as
a powerful method to probe helical edge modes.\citep{14} InAs/GaSb
material system is well suited for the task, due to its low Schottky
barrier and good interface to superconductors.\citep{15,16,17}

In this Letter, we study inverted InAs/GaSb quantum wells (QWs) contacted
by superconducting electrodes. We observe strong zero-bias dips in
the differential resistance as the Fermi level is tuned across the
hybridization gap via a front gate. Analysis of the relative size
of the dips and corresponding gap excess current is in agreement with
expectations of perfect Andreev reflection of the helical edge modes.
The perfect AR occurs in spite of a finite barrier at the interface,
with the interface transmissivity estimated to $T=0.7$. Excess current
and differential resistance dips show only a weak temperature dependence
for temperatures lower than the critical temperature of the superconducting
electrodes. On the other hand, weak magnetic fields of less than $\unit[50]{mT}$
are sufficient to completely supress excess current in the hybridization
gap, indicating strong sensitivity to time-reversal breaking.

InAs/GaSb QWs contain both electron and hole two-dimensional gases
situated in InAs and GaSb layers respectively, and enclosed with AlSb
barriers. Sample structure is shown in Fig. 1 inset a. In the inverted
regime, the electron subband is lower the than hole subband leading
to band anti-crossing and mini-gap opening.\citep{18,19,20} Energy
spectrum with the resulting hybridization gap is shown in Fig. 1 inset
b. Due to the band inversion, helical edge modes appear in the mini
gap.\citep{7,9} In order to probe the helical character of the edge
modes, superconducting niobium electrodes with critical temperature
of $T_{c}=\unit[8.27]{K}$ (BCS gap of $\Delta_{S}=\unit[1.24]{meV}$)
are deposited directly on the InAs layers, while the electrostatic
front gate is used to tune the Fermi energy $E_{F}$ into the hybridization
regime.

Andreev reflection\citep{21} is a process unique to the superconductor-normal
metal (S-N) interface, where impinging normal quasiparticle retroreflects,
having thus not only opposite velocity but also opposite charge, and
resulting in the enhancement of the total current across the interface.
The electrical current through a single S-N interface can be calculated
using the Blonder-Tinkham-Klapwijk (BTK) model:\citep{22} \begin{equation}
I=\frac{N\cdot e}{h}\int\left[f\left(E+eV\right)-f\left(E\right)\right]\left[1+A\left(E\right)-B\left(E\right)\right]dE\label{eq. 1}\end{equation}
 where $N$ is the number of modes in the normal conductor, $f\left(E\right)$
is the equilibrium Fermi distibution function, $V$ is the voltage
drop at the interface, and $A\left(E\right)$ and $B\left(E\right)$
are probabilities for Andreev and normal reflection (NR) of the electron
at the interface. In the case of ideal interface, and for biases within
the superconducting gap $\left(V<\frac{\Delta_{S}}{e}\right)$, quasi-particles
are only Andreev reflected. This is because within the superconducting
gap transmission is prohibited, and there is no potential barrier
which would absorb the momentum difference necessary for normal reflection.
In practice, due to native oxides or Schottky barriers, a potential
step always exists at the S-N interface, allowing for normal reflection
and hence reducing the probability for Andreev reflection. The interface
barrier is characterized by the scattering parameter $Z$ which is
related to the normal transmission of the barrier as $T=\frac{1}{1+Z^{2}}$.
For $Z<1$, AR dominates over NR resulting in zero bias dips in differential
resistance $dV/dI$. In this case, current enhancement due to AR also
manifests itself as an excess current $I_{excess}$, which is obtained
by extrapolating linear $I-V$ curve at high biases, i.e. for $V\gg\frac{\Delta_{S}}{e}$,
to zero bias.

\begin{figure}
\includegraphics[scale=0.9]{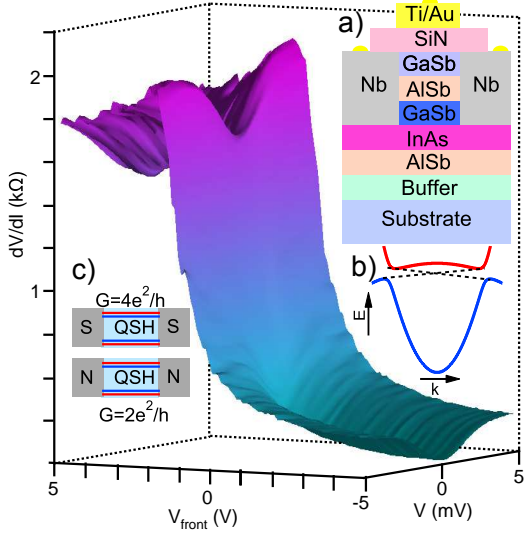}
\caption{\label{FIG. 1} {\textbf{Transport in S-QSH-S junctions.} Figure
shows differential resistance $dV/dI$ vs bias voltage $V$ across
the S-InAs/GaSb-S junction and vs front gate bias $V_{front}.$ Inset
\textbf{a} shows device cross-section while inset \textbf{b} shows
energy spectrum of inverted InAs/GaSb QWs with linearly dispersing
helical edge modes in the mini-gap. As the Fermi level $E_{F}$ is
tuned across the mini-gap via $V_{front}$, $dV/dI$ exhibits strong
peak at larger $V$. On the other hand, for $V$ close to zero, $dV/dI$
exhibits strong dips, suggesting transport dominated by Andreev reflection
processes. Inset \textbf{c} shows two-terminal structure with superconducting
and normal leads. Due to the perfect Andreev reflection at S-QSH interfaces,
voltage drop at each contact is halfed, leading to doubling of differential
conductance compared to N-QSH case. } }

\end{figure}

\begin{figure}
\includegraphics{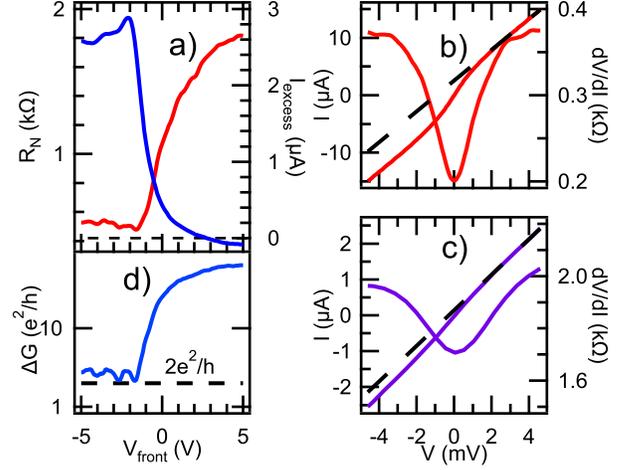}
\caption{\label{FIG. 2} {\textbf{Perfect Andreev reflection of helical edges.
}Panel \textbf{a} shows normal resistance $R_{N}$ (in blue) and excess
curent due to Andreev reflection $I_{excess}$ (in red) vs $V_{front}$.
As $V_{front}$ is decreased, $E_{F}$ is tuned towards the mini-gap
and $R_{N}$ increases towards the peak value of $\sim\unit[2]{k\Omega}$,
while concurrently $I_{excess}$ decreases from the maximal value
of $\unit[\sim2.6]{\mu A}$ ($V_{front}=\unit[5]{V}$) to mini-gap
value $I_{excess}\sim\unit[150]{nA}$ ($V_{front}=\unit[-2.1]{V}$).
Panel \textbf{b} and \textbf{c} show $dV/dI$ and $I$ vs $V$ for
$V_{front}=\unit[5]{V}$ and $V_{front}=\unit[-2.1]{V}$ respectively.
Excess current is determined as an intercept of the linear fit to
the $I-V$ curve for large $V$. Panel \textbf{d} shows conductance
difference $\Delta G\equiv G\left(V=0\right)-G\left(V\gg\Delta_{S}/e\right)$
vs $V_{front}$ on a log scale. For $E_{F}$ in the mini-gap $\Delta G$
plateaus at $2e^{2}/h$, indicating perfect AR of helical edge channels.}}

\end{figure}

Fig. 1 shows $dV/dI$ vs bias voltage $V$ across the S-InAs/GaSb-S
junction and front gate bias $V_{front}.$ As $E_{F}$ is tuned into
the mini-gap via $V_{front}$, $dV/dI$ exhibits strong peak at larger
biases $\left(V\gg\frac{\Delta_{S}}{e}\right)$. On the other hand,
for $V<\frac{2\Delta_{S}}{e}$, $dV/dI$ exhibits strong dips, i.e.
enhanced conduction due to AR. Fig. 2a shows normal resistance $R_{N}$,
i.e. $dV/dI$ for $V\gg\frac{\Delta_{S}}{e}$, vs $V_{front}$ (in
blue) and $I_{excess}$ vs. $V_{front}$ (in red). As $E_{F}$ is
tuned towards the mini-gap, $R_{N}$ increases towards the peak value
of $\sim\unit[2]{k\Omega}$ signaling mini-gap entry, while concurrently
$I_{excess}$ decreases from the maximal value of $\unit[\sim2.6]{\mu A}$
to the mini-gap value of $I_{excess}\sim\unit[150]{nA}$. In Fig.
2b we plot $I-V$ and $dV/dI-V$ curves for $V_{front}=\unit[5]{V}$.
In this case $E_{F}$ is high above the hybridization gap. Strong
zero-bias dips in $dV/dI-V$ curve are observed, while $I-V$ shows
evident non-linear character. Extrapolating current from high biases
gives $I_{excess}\sim\unit[2.6]{\mu A}$. The scattering parameter
of the barrier can be estimated from the ratio $\frac{e\cdot I_{excess}\cdot R_{N}}{\Delta_{S}}\sim0.76$,
to give $Z=0.65$ and normal transmissivity of $T=0.7$.\citep{23,24}
This transmissivity is only slightly lower than the largest reported
value of 0.86 for the InAs material system.\citep{17} In spite of
high barrier transmissivity, the absence of supercurrent in our structures
suggests that coherence is not preserved across the junction, presumably
due to the surface degradation during plasma cleaning. Nevertheless,
this simplifies the analysis in the case when $E_{F}$ is in the mini-gap,
allowing us to add conductance contributions from each S-QSH interface
independently, as previously done in N-QSH-N structures.\citep{8,9}

\begin{figure}
\includegraphics[scale=0.9]{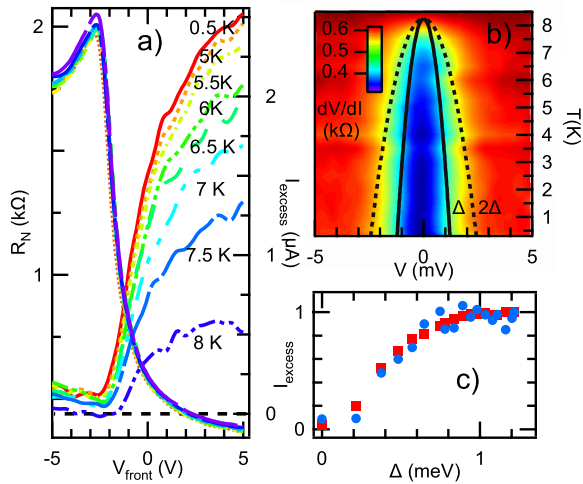}
\caption{\label{FIG. 3} {\textbf{Temperature dependence.} Panel \textbf{a}
shows $R_{N}$ and $I_{excess}$ vs $V_{front}$ for temperature $T$=$\unit[0.5]{K}$,
and $T$ from $\unit[5]{K}$ to $\unit[8]{K}$ varied in $\unit[0.5]{K}$
increments. Dependence is exceptionally weak except when $T$ approaches
$T_{c}=\unit[8.27]{K}$. Panel \textbf{b} shows color map of $dV/dI$
vs $V$ and $T$ ($V_{front}=\unit[0]{V)}$. Full and dashed lines
show BCS dependence of the superconducting gap $\Delta_{S}/e$ and
$2\Delta_{S}/e$ respectively. Dips in $dV/dI$ follow closely the
BCS gap $\Delta_{S}$. Panel \textbf{c} shows normalized $I_{excess}$,
i.e. $I_{excess}\left(T\right)/I_{excess}\left(\unit[300]{mK}\right)$,
vs $\Delta_{S}\left(T\right)$ for $E_{F}$ above the mini-gap (in
red) and $E_{F}$ in the mini-gap (in blue). In both cases, normalized
$I_{excess}$ shows equal decrease as the $\Delta_{S}$ is reduced
with $T$. } }

\end{figure}

In the case of S-QSH single edge interface, the absence of back-scattering
channels in the helical edge requires NR probability $B\left(E\right)=0$
at all energies. Within the superconducting gap $\left(E<\Delta_{S}\right)$,
electron transmission is excluded, requiring a perfect AR with probability
$A\left(E\right)=1$.\citep{14} Evaluating equation \eqref{eq. 1}
in zero temperature limit for this case gives a contact resistance
for a single helical edge channel of $\frac{h}{4e^{2}}$ when $V<\frac{\Delta_{S}}{e}$.
In two-terminal geometry, used in our experiments, this gives a resistance
of each helical edge mode to be $\frac{h}{4e^{2}}$+$\frac{h}{4e^{2}}$=$\frac{h}{2e^{2}},$
giving a total two-terminal resistance of $\frac{h}{2e^{2}}||\frac{h}{2e^{2}}=\frac{h}{4e^{2}}$.
On the other hand, for $V\gg\frac{\Delta_{S}}{e}$, electron transmission
into the superconducting lead becomes possible and AR probability
scales to zero as $A\left(E\right)\backsim\left(\frac{\Delta_{S}}{E}\right)^{2}\rightarrow0$,\citep{22}
reducing equation \eqref{eq. 1} to the familiar case of N-QSH interface
with a contact resistance of $\frac{h}{2e^{2}}$. Simple resistance
combination now gives a total two-terminal resistance of $\frac{h}{2e^{2}}$.

In InAs/GaSb QWs this analysis may be further complicated by the presence
of low mobility mini-gap bulk carriers.\citep{20,25} However, we
note here that scattered states, which lead to residual bulk conductivity
due to their loss of quantum properties and inability to tunnel,\citep{20,25,26}
are not expected to participate in AR which is a quantum process.
As a result, difference between two-terminal conductances at zero
and high biases will be: $\Delta G\equiv G\left(V=0\right)-G\left(V\gg\Delta_{S}/e\right)=\frac{2e^{2}}{h}$(Fig.
1 inset c). Note that in Fig. 2c, where $E_{F}$ is in the hybridization
gap, $\frac{dV}{dI}\left(V=0\right)\sim\unit[1.7]{k\Omega,}$ while
$\frac{dV}{dI}\left(V\gg\Delta_{S}/e\right)\sim\unit[2]{k\Omega.}$
Inverting these two values gives $\Delta G\sim2.2\frac{e^{2}}{h}$,
which is suprisingly close to the expected value of $\frac{2e^{2}}{h}$.
This is better illustrated in Fig. 2d, which shows plateauing of $\Delta G$
to a conductance value of $\frac{2e^{2}}{h}$, as $E_{F}$ is pushed
into the hybridization gap, validating the picture of perfect AR of
helical edge channels. 

\begin{figure}
\includegraphics[scale=1.1]{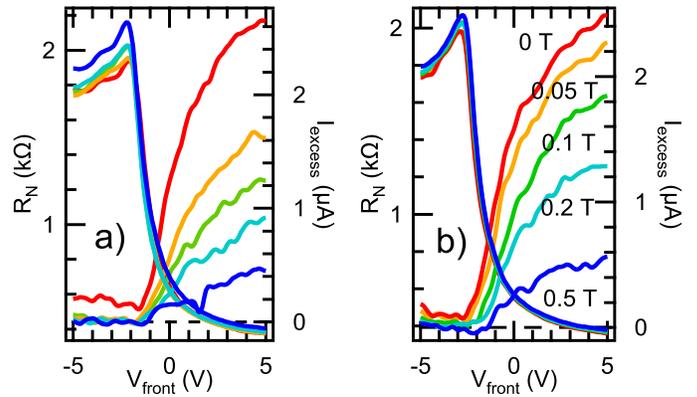}
\caption{\label{FIG. 4} {\textbf{Magnetic field dependence.} Panels show
$R_{N}$ and $I_{excess}$ vs $V_{front}$ for perpendicular magnetic
fields of $B_{\bot}=\unit[0]{T},$ $\unit[0.05]{T}$, $\unit[0.1]{T},$
$\unit[0.2]{T},$ and $\unit[0.5]{T}$ in \textbf{a} and in \textbf{b}
for in-plane magnetic fields $B_{||}$ with the same increments. Although
for $E_{F}$ above the hybridization gap, $I_{excess}$ survives up
to $\unit[0.5]{T}$, for $E_{F}$ in the mini-gap $I_{excess}$ is
completely supressed with $B_{\bot}=\unit[0.05]{T}$ and $B_{||}=\unit[0.1]{T}$.
This is in contrast to the equal supression of $I_{excess}$ in temperature
dependence (Fig. 3c), suggesting different nature of excess current
in and outside of the hybridization gap. } }

\end{figure}

Furthermore, dissimilar conductance values within and outside of the
superconductive gap translate into non-linear $I-V$ curve which can
be approximated by $I=\left(\frac{4e^{2}}{h}+G_{bulk}\right)V$ for
$V<\frac{\Delta_{S}}{e}$ and $I=\left(\frac{2e^{2}}{h}+G_{bulk}\right)V+\frac{2e\Delta_{S}}{h}$
for $V>\frac{\Delta_{S}}{e}$. The intercept of the latter equation
gives an estimate of the excess current as $I_{excess}\sim\frac{2e\Delta_{S}}{h}\sim\unit[100]{nA}$.
Considering the approximative character of the given analysis, the
latter value is in reasonable agreement with the measured value of
$\unit[150]{nA}$ in Fig. 2a and 2c.

The temperature dependence of $I_{excess}$ in Fig. 3a shows only
a weak dependence for temperatures up to $\unit[6.5]{K}$ and it is
quickly supressed as the temperature is further increased towards
the critical temperature of niobium leads. Furthermore, a color map
of temperature evolution of $dV/dI$ is shown in Fig. 3b, with dips
in $dV/dI$ closely following the BCS temperature dependence of superconducting
gap $\Delta_{S}$. We note here that $I_{excess}$ for $E_{F}$, both
inside and outside of the mini-gap, show comparative supression when
$\Delta_{S}$ is reduced with increased temperature. This is most
easily seen when $I_{excess}$ is normalized by the corresponding
low temperature values, i.e. $I_{excess}\left(T\right)/I_{excess}\left(\unit[300]{mK}\right)$
and plotted in Fig. 3c for these two cases. 

This is in sharp contrast to the magnetic field dependence of $I_{excess}$
shown in Fig. 4, where $I_{excess}$ for $E_{F}$ in the mini-gap
is supressed much faster than in the case when $E_{F}$ is outside
of the mini-gap. In fact, perpendicular magnetic fields of less than
$\unit[50]{mT}$ are sufficient to fully supress AR processes in the
mini-gap, while above the mini-gap AR processes survive in fields
up to at least $\unit[500]{mT}$. Similar disproportionality is also
observed for the in-plane magnetic fields, albeit in this case mini-gap
$I_{excess}$ survives for fields up to $\unit[100]{mT}$ while above
the mini-gap, AR processes are still observable at $\unit[500]{mT}$.
Such sensitivity to time-reversal breaking indeed suggests that the
observed mini-gap $I_{excess}$ is due to the perfect AR of helical
edge modes. Applying small magnetic fields destroys the perfect destructive
interference of back-scattering paths,\citep{2} opening the back-scattering
channels in our structures. In this case, the probability of AR decreases,
and $I_{excess}$ vanishes.

In conclusion, we probe the recently discovered helical edge modes
in InAs/GaSb QWs via Andreev reflection. Strong zero-bias dips in
the differential resistance are observed as the Fermi level is tuned
across the mini-gap. Evolution of the mini-gap differential resistance
with applied bias as well as measured mini-gap excess current of $\sim\unit[150]{nA}$
are in good agreement with the prediction of perfect Andreev reflection
of the helical edge modes, necessitated by the absence of back-scattering
channels. The perfect AR occurs in spite of a finite barrier at the
interface and shows strong sensitivity to time-reversal breaking -
hallmarks of helical nature of the QSH edges. Although sufficient
coherence is not achieved in the junctions to observe a supercurrent,
with further optimization in fabrication, InAs/GaSb readily arises
as a viable platform where theoretical predictions of Majorana fermion
modes\citep{11,12,13} can be experimentally explored. 

The work at Rice was supported by Rice Faculty Initiative Fund, Hackerman
Advanced Research Program grant 003604-0062-2009, Welch Foundation
grant C-1682, and NSF grant DMR-0706634. I.K. acknowledges partial
support from M. W. Keck Scholar. We thank D. C. Tsui for insightful
discussions.

\textcompwordmark{}

\textbf{Materials and Methods}

The experiments are performed on high quality 125Å InAs/50Å GaSb quantum
wells in the inverted regime, patterned in a superconductor-normal
metal-superconductor (S-N-S) junction geometry. Superconducting niobium
electrodes with critical temperature of $T_{c}=\unit[8.27]{K}$ are
deposited directly on InAs layers via magnetron sputtering. Top layers
of the contact region are selectively removed by etching, and plasma
cleaned in argon atmosphere in-situ prior to niobium deposition.\citep{15,16,17}
The width and length of the junctions are $W\sim\unit[1]{\mu m}$
and $L\sim\unit[0.5]{\mu m}$. The front gate is fabricated by depositing
SiN using plasma enhanced chemical vapor deposition system, and evaporating
a Ti/Au metal gate. Additional sample and processing details are given
elsewhere.\citep{9,20}

\textcompwordmark{}

\textbf{Author contributions}

I.K. fabricated the devices, performed the measurements, and analysed
the data. G. S. prepared the sample wafer. R.R.D. supervised and provided
continuous guidance for the experiments and analysis. Manuscript was
prepared by I. K. and R. R. D.

\textcompwordmark{}

The authors declare no competing financial interests.


\begin{thebibliography}{27}
\bibitem[27]{27}Recent experiments\citep{9} show that low bulk conduction
has little effect on edge conduction, presumably due to the large
disparity in Fermi wave-vectors between bulk and edge states, allowing
for simple addition of edge and bulk contributions.
\end{thebibliography}
\end{document}